\documentclass[journal]{IEEEtran}

\ifCLASSINFOpdf
\else
   \usepackage[dvips]{graphicx}
\fi
\usepackage{url}

\usepackage{graphicx}
\usepackage{bm}
\usepackage{mathtools}
\usepackage{amsfonts}
\usepackage{amssymb}	
\usepackage{hyperref}
\hypersetup{breaklinks,colorlinks,citecolor=blue}
\usepackage{breakurl}
\usepackage{booktabs}
\usepackage{graphicx}
\usepackage{wrapfig}
\usepackage[SF]{subfigure}
\usepackage{xspace}
\usepackage{eurosym}
\usepackage{longtable}

\usepackage{tikz}
\tikzset{
  mylabel/.style = {font=\footnotesize, midway, fill=white, anchor=center}
}

\DeclareMathOperator*{\argminT}{argmin}
\DeclareMathOperator*{\argmaxT}{argmax}

\usepackage{mathtools}
\DeclarePairedDelimiterX{\norm}[1]{\lVert}{\rVert}{#1}
\usepackage{scalerel}
\newcommand\ta{\scalerel*{\tau}{T}}
\newcommand\R{\mathcal{R}}
\newcommand\Ll{\mathcal{L}}
\newcommand{\SOT}{\text{SO}(3)}
\newcommand{\St}{\mathbb{S}^2}
\newcommand{\Bt}{\mathbb{B}^3}
\newcommand{\Bf}{\mathbb{H}^4}
\newcommand{\PHI}{\bm{\mathsf{\Phi}}}
\newcommand{\Rp}{\mathbb{R}^+}

\newcommand*\dif{\mathop{}\!\mathrm{d}}
\DeclarePairedDelimiterX{\bnorm}[1]{{}_{\mathbb{B}^3}\!\lVert}{\rVert}{#1}
\DeclarePairedDelimiterX{\snorm}[1]{{}_{\mathbb{S}^2}\!\lVert}{\rVert}{#1}
\DeclarePairedDelimiterX{\pnorm}[1]{{}_{\Psi}\!\lVert}{\rVert}{#1}

\begin{document}

\title{Bayesian variational regularization on the ball}

\author{Matthew A. Price and Jason D. McEwen
  \thanks{MAP is supported by the Science and Technology Facilities Council (STFC).
    This work is supported in part by the Leverhulme Trust.}
  \thanks{E-mail: m.price.17@ucl.ac.uk}}

\markboth{IEEE Signal Processing Letters, Submitted}
{Shell \MakeLowercase{\textit{et al.}}: Bare Demo of IEEEtran.cls for IEEE Journals}
\maketitle

\begin{abstract}
  We develop variational regularization methods which leverage sparsity-promoting priors to solve severely
  ill-posed inverse problems defined on the 3D ball (\emph{i.e.} the solid sphere).
  Our method solves the problem natively on the ball and thus does not suffer from discontinuities that plague alternate approaches where each spherical shell is considered independently.
  Additionally, we leverage advances in probability density theory to produce Bayesian variational methods which benefit from the computational efficiency of advanced convex optimization algorithms,
  whilst supporting principled uncertainty quantification. We showcase these variational regularization and uncertainty quantification techniques on an illustrative example. The C++ code
  discussed throughout is provided under a GNU general public license.
\end{abstract}


\IEEEpeerreviewmaketitle

\section{Introduction}
\label{sec:introduction}

\IEEEPARstart{I}{nverse} problems on Euclidean manifolds have been researched extensively and associated techniques have found effective application in countless domains. However,
increasingly often one wishes to consider inverse problems defined on curved, non-Euclidean
manifolds, \emph{e.g.} diffusion magnetic resonance imaging (MRI) \cite{tuch:2004} and 2D dark matter reconstructions on
the sphere, and many aspects of geophysics \cite{simons:2011, marignier:2020, kendall:2021},
astrophysics \cite{heavens:2003, leistedt:2015}, and molecular modeling \cite{boomsma:2017}
on the 3D ball, for which very few techniques have been developed.

Inverse problems are often solved by Bayesian Markov chain Monte Carlo (MCMC) sampling methods or variational approaches
(optimization \emph{etc.}). MCMC methods
are highly computationally demanding on the ball, due to the computational complexity of transforms on curved manifolds, and are infeasible for many applications.
Variational methods, which solve inverse problems through classical optimization
techniques, are typically scalable and robust, supporting both theoretical guarantees, and can support principled uncertainty quantification (as demonstrated in this letter).
Such techniques are thus perfectly suited to scientific analysis on the ball, where computational efficiency and probabilistic interpretations are highly desirable. Variational methods have been
considered over the sphere \cite{mcewen:2013:sparse, wallis:2017:sparse, price:2021:s2inv},
often leveraging ideas from compressed sensing \cite{donoho:2006:compressed,
  candes:2006:compressive}, and typically promoting sparsity in spherical wavelet dictionaries, \emph{e.g.}
\cite{chan:2017:curvelets, leistedt:2013:s2let, mcewen:2019:ridgelets}, to recover state-of-the-art
results. Spherical techniques have been used tomographically (as concentric spherical shells) to model
radially distributed datasets, however holistic approaches, which perform analysis natively on
the underlying manifold (the ball), are crucially missing. Wavelet transforms on the 3D ball have been
developed to support radially distributed problems \cite{michel:2005, fengler:2006,lanusse:2012, durastanti:2014,
khalid:2016, leistedt:2012}, however these dictionaries have, to our best knowledge, not been
leveraged to perform variational inference on the ball.

In this letter we develop scalable techniques, with associated open-source software,
which leverage variational regularization methods to solve ill-posed and/or ill-conditioned
inverse problems natively on the ball. Furthermore, leveraging recent developments in the
theory of probability density theory \cite{pereyra:2017}, we demonstrate how convex variational regularization
techniques can be combined with advances in probability density theory to construct computationally efficient 
signal reconstruction techniques on the ball with principled uncertainty quantification, or `Bayesian variational regularization'.

\section{Bayesian variational regularization on ball} \label{sec:core}

In this section we develop mathematical techniques for the analysis of spin signals on the 3D ball and wavelets
on the directional ball, scalable convex optimization algorithms on the ball,
and variational regularization techniques which support principled Bayesian uncertainty quantification. Throughout
we adopt separable eigenfunctions on the ball, with radial basis functions given by the
Laguerre polynomials \cite{pollard:1947, weniger:2008} and angular basis functions given by
the spin spherical harmonics $_sY_{\ell m}$ \cite{newman:1966, goldberg:1967, mcewen:2011:ssht, mcewen:2015:s2let}.
As spin spherical harmonic transform are more common in the associated literature, we will focus primarily on the novel radial components \cite{leistedt:2012},
and the Bayesian interpretation \cite{pereyra:2017, cai:2018:uq, price:2018, price:2021:darkmapper}.

\subsection{Spin signals on the ball} \label{sec:spin_signals}
Here we discuss the construction of Spherical-Laguerre basis functions on the 3D ball, developed in previous work \cite{leistedt:2012} and adopted throughout this letter. First let us define the Laguerre basis functions along the radial half-line $K_p(r)$ as
\begin{equation} \label{eq:spherical_laguerre_basis}
  K_p(r) \equiv \sqrt{\frac{p!}{(p+2)!}} \frac{e^{\frac{-r}{2\tau}}}{\sqrt{\tau^3}} L_p^{(2)}(\frac{r}{\tau}),
\end{equation}
where $L_p^{(2)}$ is the $p^{\text{th}}$-associated $2^{\text{nd}}$-order Laguerre polynomial \cite{pollard:1947, weniger:2008},
and \mbox{$\tau \in \Rp$} is a scale factor that adds a scaling flexibility. These basis
functions are orthonormal on $\Rp$, \emph{i.e.} \mbox{$\langle K_p | K_q \rangle_{\Rp} = \delta_{pq}$},
and complete, by Gram-Schmidt orthogonalization and exploiting
polynomial completeness on \mbox{$\mathsf{L}^2(\Rp, r^2 e^{-r}\dif r)$} \cite{leistedt:2012}. Any square-integrable function \mbox{$f \in \mathsf{L}^2(\Rp)$} can be projected into this basis as
\begin{equation} \label{eq:forward_projection}
  f_p  = \langle f | K_p \rangle = \int_{\Rp} \dif r r^2 f(r) K_p(r),
\end{equation}
which supports exact synthesis by
\begin{equation} \label{eq:inverse_projection}
  f(r) = \sum_{p=0}^\infty f_p K_p(r).
\end{equation}
Real-world functions are typically to a good approximation bandlimited, \emph{i.e.} the Fourier-Laguerre coefficients of signals \mbox{$f \in \Rp$} are such that
\mbox{$f_p = 0, \forall p \geq P$}, and so this summation is truncated at $P$. We adopt the Gauss-Laguerre quadrature (see \emph{e.g.} \cite{press:2007}), which is commonly used to numerically evaluate
integrals over the radial half-line, and was used to develop an exact sampling theorem on Spherical-Laguerre space \cite{leistedt:2012}.

Suppose we adopt these radial basis functions which we then combine with the spin-$s$ spherical harmonic angular basis functions \mbox{$_sY_{\ell m}(\omega)$} \cite{newman:1966, goldberg:1967} for \mbox{$s \in \mathbb{Z}$} and \mbox{$\omega = (\theta, \psi) \in \St$}, where $\theta = [0, \pi)$ is the colatitude and $\psi \in [0, 2\pi)$ is the longitude. In such a case, we can straightforwardly define the
Spherical-Laguerre basis
functions \mbox{$_sZ_{\ell m p}(\bm{r}) = K_p(r) _sY_{\ell m}(\omega)$} for \mbox{$\bm{r} = (r, \omega) \in \Bt \coloneqq \Rp \times \St$}, which are orthogonal \mbox{$\langle _sZ_{\ell m p} | _sZ_{\ell^\prime m^\prime p^\prime} \rangle_{\Bt} = \delta_{\ell \ell^\prime}\delta_{m m^\prime}\delta_{p p^\prime}$} and onto which any square integrable spin-$s$ function on the ball \mbox{$_sf \in \mathsf{L}^2(\Bt)$} can
be projected by
\begin{equation}
  _sf_{\ell m p} = \langle _sf | _sZ_{\ell m p} \rangle_{\Bt} = \int_{\Bt} \dif \mu(\bm{r}) _sf(\bm{r}) _sZ_{\ell m p}^*(\bm{r}),
\end{equation}
where \mbox{$\dif \mu(\bm{r}) = \dif^3\bm{r} = r^2\sin{\theta} \dif r \dif \theta \dif \psi$} is the rotation invariant measure (Haar measure) on the ball. By considering the separability and completeness
of angular and radial basis functions this projection supports exact synthesis, such that
\begin{equation}
  _sf(\bm{r}) = \sum_{p=0}^{P-1} \sum_{\ell=0}^{L-1} \sum_{m=-\ell}^{\ell} {}_sf_{\ell m p} {}_sZ_{\ell m p} (\bm{r}),
\end{equation}
where $L, P \in \mathbb{Z}^+$ are the angular \cite{mcewen:2011:ssht} and
radial \cite{leistedt:2012} bandlimits respectively. In this work, by considering the relations 
presented in this subsection, fast adjoint Spherical-Laguerre transforms were constructed, facilitating 
variational regularization on the ball (see Section \ref{sec:examples}).

\subsection{Directional scale-discretized wavelets on the ball} \label{sec:ball_wavelets}

Here we extend the Spherical-Laguerre wavelets on the 3D ball developed in previous work \cite{leistedt:2012} to 4D directional scale-discretized wavelets on the ball. Furthermore, we extend the discussion to include spin-signals, which arise in various areas of physics \emph{e.g.} quantum mechanics and weak gravitational lensing \cite{price:2021:darkmapper}.
Consider the radial translation operator $\ta_r$ for \mbox{$r\in\Rp$} (see \cite{leistedt:2012, mcewen:2013:flaglet} for further details), and rotation $\R_\rho$, for Euler angles
\mbox{$\rho = (\alpha, \beta, \gamma) \in \SOT$} with \mbox{$\alpha \in [0, 2\pi)$}, \mbox{$\beta \in [0, \pi)$},
and \mbox{$\gamma \in [0, 2\pi)$}, with action
\mbox{$(\R_\rho {}_s f )(\omega) \equiv e^{-is\theta} {}_s f (\R_\rho^{-1} \bm{\omega})$}. Further define the
concatenation of these transforms to be the 4D transformation
\mbox{$\Ll_{\bm{h}} = \ta_r \R_\rho$ for $\bm{h} = (r, \rho) \in \Bf \coloneqq \Rp \times \SOT$}. Leveraging this composite transformation one
can straightforwardly define the directional wavelet coefficients $W^{{}_s\Psi^{j j^{\prime}}} \in \mathsf{L}^2(\Bf)$ of any square integrable
spin-$s$ function \mbox{${}_sf \in \mathsf{L}^2[\Bt]$} by the directional convolution $\circledast$
\begin{align} \label{eq:directional_convolution}
  W^{{}_s\Psi^{j j^{\prime}}} (\bm{h}) & \equiv ( {}_sf \circledast {}_s\Psi^{j j^{\prime}} ) (\bm{h}) \equiv \langle {}_sf, \Ll_{\bm{h}} {}_s\Psi^{j j^{\prime}} \rangle_{\Bt} \nonumber \\
                                       & = \int_{\Bt} \dif \mu(\bm{r})
  {}_sf(\bm{r}) (\Ll_{\bm{h}} {}_s\Psi^{j j^{\prime}})^\star (\bm{r}),
\end{align}
where \mbox{${}_s\Psi^{j j^{\prime}} \in \mathsf{L}^2[\Bt]$} is the wavelet kernel at angular and radial scales \mbox{$j, j^{\prime} \in \mathbb{Z}^+$} respectively. These scales determine the volume over which a given wavelet function has compact support \cite{leistedt:2012}.
Typically, wavelet coefficients do not capture low frequency signal content, which instead is
captured by axisymmetric scaling functions \mbox{${}_s \Upsilon \in \mathsf{L}^2(\Bt)$} with coefficients
\mbox{$W^{{}_s\Upsilon} \in \mathsf{L}^2(\Bt)$} defined by the axisymmetric convolution
$\odot$ with a spin-$s$ signal \mbox{${}_sf \in \mathsf{L}^2(\Bt)$} such that
\begin{align}
  W^{{}_s\Upsilon} (\bm{r}) & \equiv ({}_sf \odot {}_s\Upsilon)(\bm{r}) \equiv \langle {}_sf, \mathcal{L}_{\bm{r}} {}_s\Upsilon \rangle_{\Bt} \nonumber \\
                            & = \int_{\Bt}
  \dif \mu(\bm{r}^\prime) {}_sf(\bm{r}^\prime)(\Ll_{\bm{r}} {}_s\Upsilon)^\star
  (\bm{r}^\prime)\, ,
\end{align}
where $\Ll_{\bm{r}}$ is an axisymmetric simplification of the full 4D transformation
$\Ll_{\bm{h}}$. For suitable choices of wavelet and scaling generating functions (those which satisfy wavelet admissibility) these
projections support exact synthesis by
\begin{align}
  _sf(\bm{r}) & = \int_{\Bt} \dif \mu(\bm{r}^\prime) W^{{}_s\Upsilon} (\bm{r}^\prime)(\Ll_{\bm{r}^\prime} {}_s\Upsilon)
  (\bm{r}) \nonumber                                                                                                                                                                     \\
  +           & \sum_{j=J_0}^J \sum_{j^\prime=J^\prime_0}^{J^\prime} \int_{\Bf} \dif \mu(\bm{h}) W^{{}_s\Psi^{j j^\prime}} (\bm{h}) (\Ll_{\bm{h}} {}_s\Psi^{j j^\prime}) (\bm{r}),
\end{align}
where $\dif \mu(\bm{h}) = \dif^4\bm{h} = r^2 \sin{\beta} \dif r \dif \alpha \dif \beta \dif \gamma$ is the Haar measure on $\Bf$. By construction \cite{leistedt:2012} this wavelet dictionary exhibits both good frequency and
spatial localization, permits exact synthesis, and leverages optimal sampling theories for
efficient transforms. Furthermore, by adopting adjoint Spherical-Laguerre transforms (see subsection \ref{sec:spin_signals}) 
fast adjoint 4D wavelet transforms on the ball were constructed.

\subsection{Efficient transformations on the ball} \label{sec:computational_scaling}

Variational methods on the ball require an additional level of complexity over those defined
on the spherical manifolds, which are already significantly computationally expensive. The forward and inverse Spherical-Laguerre transforms are computed through
the FLAG\footnote{\url{https://astro-informatics.github.io/flag/}} package \cite{leistedt:2012} with computational complexity $\sim \mathcal{O}(L^4)$, built on spin spherical harmonic transforms provided by the SSHT\footnote{\url{https://astro-informatics.github.io/ssht/}} package
\cite{mcewen:2011:ssht, mcewen:2013:sparse}. Similarly, forward and inverse wavelets
transforms on the 3D ball are computed through the FLAGLET\footnote{\url{http://astro-informatics.github.io/flaglet/}} package
\cite{leistedt:2012} with computational complexity $\sim \mathcal{O}(NL^4)$, where $N \in \mathbb{Z}^+$ is the wavelet directionality, built on the wavelet transforms provided by the
S2LET\footnote{\url{https://astro-informatics.github.io/s2let/}} package \cite{wiaux:2008, leistedt:2013:s2let, mcewen:2015:s2let,
  mcewen:2019:ridgelets, chan:2017:curvelets, mcewen:2018:localisation}. Both these transforms on the ball (FLAG and FLAGLET) scale at least quartically with bandlimit. Therefore, even optimally sampled transforms on the ball are very computationally expensive, motivating attention to scalable implementations.

\subsection{Bayesian variational regularization on the ball} \label{sec:bayesian_optimisation}

Consider measurements \mbox{$\bm{y} \in \mathbb{R}^{M}$}, \emph{e.g.} observations on the
sky with some radial component, which may be related to some intrinsic underlying field of
interest on the ball \mbox{$\bm{x} \in \mathbb{R}^{N_{\Bt}}$} by a sensing operator
\mbox{$\PHI \in \mathbb{R}^{M \times N_{\Bt}} : \bm{x} \mapsto \bm{y}$}. Further suppose measurements are polluted
with noise $\bm{n}$, then our measurement model is generally given by $\bm{y} = \PHI \bm{x} + \bm{n}$,
which is both classically ill-posed in the sense of Hadamard \cite{hadamard:1902} and may be seriously ill-conditioned. There are many methods for inferring $\bm{x}$ from $\bm{y}$, in this work we will consider a Bayesian variational approach, so as to benefit from the computational
efficiency of variational methods (a key component on the ball) whilst retaining the
principled statistical interpretation provided by Bayesian methods.

In a Bayesian sense, given a sufficient understanding of our physical system (including \emph{e.g.} the forward model and the noise distribution \emph{etc.}) we can assign
a likelihood distribution \mbox{$P(\bm{y} | \bm{x}; \PHI)$}, which acts as a data-fidelity constraint on our
solutions. Furthermore, suppose we have some \emph{a priori} knowledge as to the nature of
our latent variable, \emph{e.g.} $\bm{x}$ is presumed to be sparse in a given dictionary,
then we can straightforwardly define a Bayesian prior distribution \mbox{$P(\bm{x})$}, which
acts as a regularization functional to stabilize our inference. With these distributions defined we
can construct our posterior distribution through Bayes' theorem
\begin{equation} \label{eq:bayes_theorem}
  P(\bm{x}|\bm{y}; \PHI) \propto P(\bm{y}|\bm{x}; \PHI) P(\bm{x}),
\end{equation}
where we drop the normalization term (Bayesian evidence) as it does not
affect our solution, and for simplicity. A reasonable choice of solution, in a Bayesian sense, is that which
maximizes the posterior odds (\emph{i.e.} the most likely one), called the \emph{maximum a posteriori} (MAP)
solution, given by
\begin{align}
  \bm{x}^{\text{MAP}} & \equiv \argmaxT_{\bm{x}} \big \lbrace P(\bm{x}|\bm{y};\PHI)
  \big \rbrace, \nonumber                                                                                                                                 \\
                      & \propto \argminT_{\bm{x}} \big \lbrace -\log ( \; P(\bm{y}|\bm{x};\PHI)P(\bm{x})
  \;) \big \rbrace, \nonumber                                                                                                                             \\
                      & \propto \argminT_{\bm{x}} \big \lbrace h(\bm{x}) = f(\bm{x}) + g(\bm{x}) \big \rbrace,
\end{align}
where the second line comes from the monotonicity of the logarithm function. The final
line highlights that MAP estimation, for the common class of log-concave distributions, yields convex objectives $h(\bm{x})$, and this is equivalent
to unconstrained convex optimization.
Such optimization problems typically leverage $1^{\text{st}}$-order information to efficiently converge to global (from convexity) extremal solutions \cite{combettes:2011}.
For convex but non-differentiable objectives (\emph{e.g.} sparsity priors) gradient
information is accessed through the proximal projection \cite{moreau:1962}, and thus
extremal solutions are efficiently recovered \emph{via} proximal optimization algorithms \cite{boyd:2011:admm,combettes:2011}. Such algorithms permit strong guarantees of both convergence and rate of
convergence, however they still only recover point estimates and do not naively support uncertainty quantification.

\subsection{Uncertainty quantification of MAP estimation} \label{sec:UQ}

Bayesian methods often consider credible regions (regions of high probability concentration)
\mbox{$C_{\alpha} \subset \mathbb{C}^{N_{\Bt}}$} of the full posterior distribution,
at \mbox{$100(1-\alpha)\%$} confidence, by evaluating
\begin{equation} \label{eq:credible_region_integral}
  P(\bm{x} \in C_{\alpha}|\bm{y}; \PHI) = \int_{\bm{x} \in \mathbb{R}^{N_{\Bt}}} P(\bm{x}|
  \bm{y}; \PHI)\mathbb{I}_{C_{\alpha}}d\bm{x} = 1 - \alpha,
\end{equation}
which is computationally intractable in high dimensional settings, such as data on the ball, even for moderate resolutions. In our method we adopt a recently derived conservative approximation (which is valid for all log-concave posteriors or convex objectives) to the highest
posterior density (HPD) credible set $C^{\prime}_{\alpha} \supseteq C_{\alpha}$ \cite{pereyra:2017} defined by
\begin{align}
  C^{\prime}_{\alpha} \subset \mathbb{C}^{N_{\Bt}} & \coloneqq
  \Big \lbrace \bm{x} : h(\bm{x}) \leq \epsilon^{\prime}_{\alpha} \Big \rbrace, \nonumber                    \\
  \epsilon^{\prime}_{\alpha} = h(\bm{x}^{\text{MAP}}) +                  & \sqrt{16 N \log(3 / \alpha)} + N,
  \label{eq:approximate_HPD}
\end{align}
which allows one to approximate $C_\alpha$ with knowledge only of the MAP solution
$\bm{x}^{\text{MAP}}$ and the dimension $N_{\Bt}$. This is a crucial realization for
variational methods on complex manifolds (such as the ball), as the necessity for scalable,
computationally efficient approaches is paramount. Furthermore, the approximation
error is bounded above \cite{pereyra:2017} thus affording sensitivity guarantees (\emph{i.e.} $\epsilon^{\prime}_{\alpha}$ cannot become arbitrarily larger than $\epsilon_{\alpha}$). The error in this approximation has been assessed in a variety of
application domains \cite{cai:2018:uq, price:2019:lci} and has been benchmarked against
proximal MCMC methods \cite{pereyra:2016:proximal}.

A number of uncertainty quantification techniques have recently been developed which are built
around this approximation, in a variety of settings, many of which exploit linearity \cite{price:2021:s2inv}
to facilitate extremely rapid computation.
In this letter we consider, for the first time on the ball, perhaps the most straightforward uncertainty quantification technique, Bayesian hypothesis testing \cite{cai:2018:uq, price:2018,repetti:2018}.
Bayesian hypothesis testing is conducted as follows. A feature of $\bm{x}^{\text{MAP}}$ is
adjusted to construct a surrogate solution $\bm{x}^{\text{SUR}}$ from which it is determined
if this solution belongs to the credible set at confidence $100(1-\alpha)\%$. If
$\bm{x}^{\text{SUR}}$ does not belong to $C_{\alpha}^{\prime}$ then it necessarily
does not belong to $C_{\alpha}$ (from the conservative nature of the approximation in
Equation \ref{eq:approximate_HPD}) and therefore the feature is
statistically significant at $100(1-\alpha)\%$ confidence. Conversely, if $\bm{x}^{\text{SUR}} \in C_{\alpha}^{\prime}$
then the statistical significance of the feature of interest is indeterminate. In this letter we consider features
$\Omega \subset \bm{x}^{\text{MAP}}$ to be local sub-structure and thus hypothesis
tests in this case relate to the physicality of local structure, \emph{i.e.} whether these structures
are aberrations or physical signals.

One can straightforwardly leverage Bayesian hypothesis testing to constrain the maximum and minimum intensities
a parition of $\bm{x}^{\text{MAP}}$ can take, such that the resulting surrogate
$\bm{x}^{\text{SUR}}$ saturates the approximate level-set threshold $\epsilon^{\prime}_{\alpha}$. In this sense
one can recover local voxel level Bayesian error bars coined \emph{local credible intervals} \cite{cai:2018:uq,
price:2018,repetti:2018,price:2021:s2inv}. The concept of Bayesian hypothesis testing can further be leveraged to consider hypothesis tests which quantify the uncertainty in \emph{e.g.} feature location \cite{price:2019:peaks} and global features \cite{price:2021:darkmapper}.

\begin{figure}
\centering
  \begin{minipage}{0.38\linewidth}
    \centering \includegraphics[width=\linewidth, trim={2.9cm 0.7cm 2.9cm 0cm}, clip]{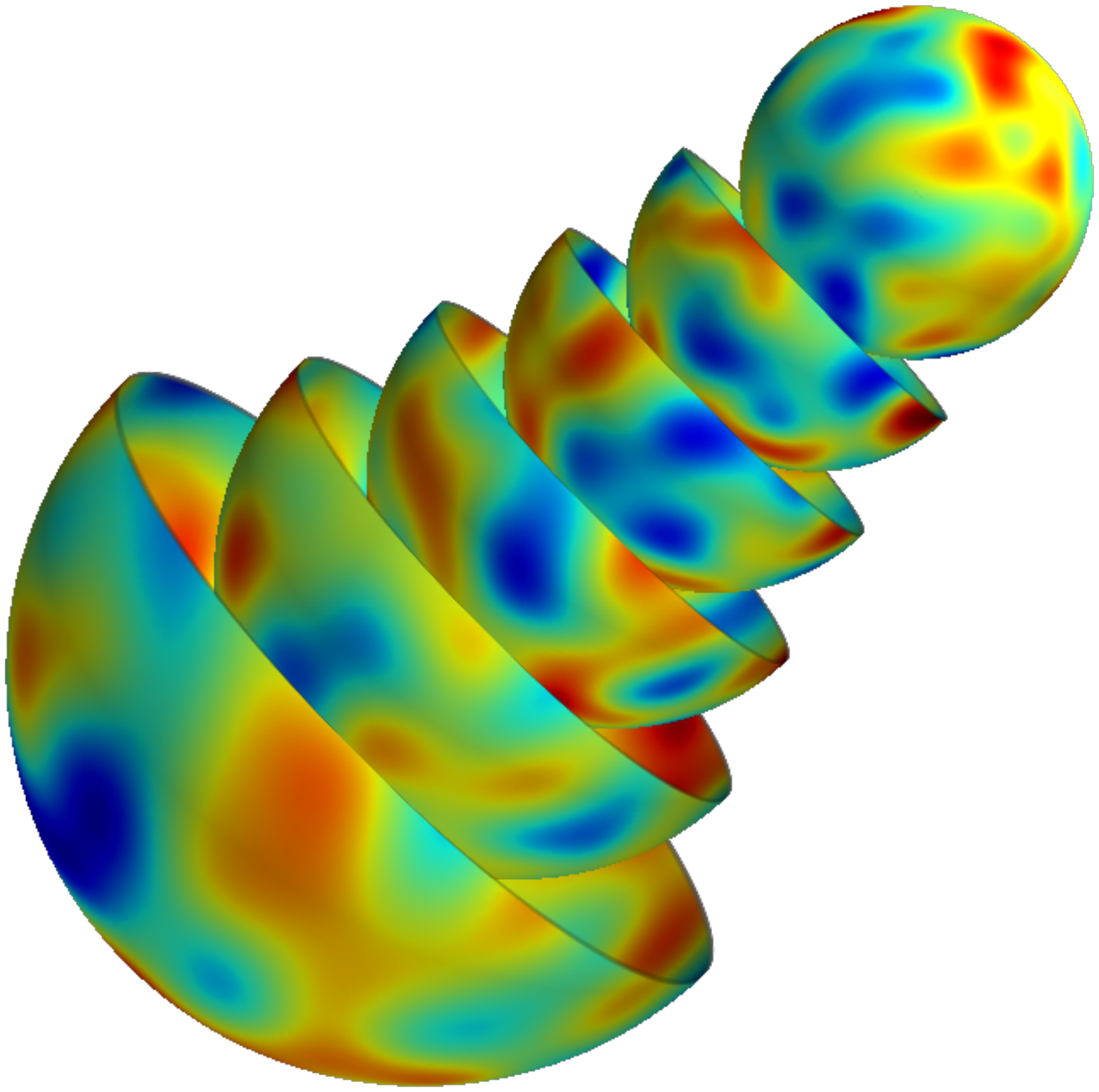}
    \put(-72,64){$\bm{x}^{\text{TRUE}}$}
    \put(-92, -3){\footnotesize}
  \end{minipage}
  \begin{minipage}{0.38\linewidth}
    \centering \scalebox{-1}[1]{\includegraphics[width=\linewidth, trim={2.3cm 0.7cm 2.3cm 0cm}, clip]{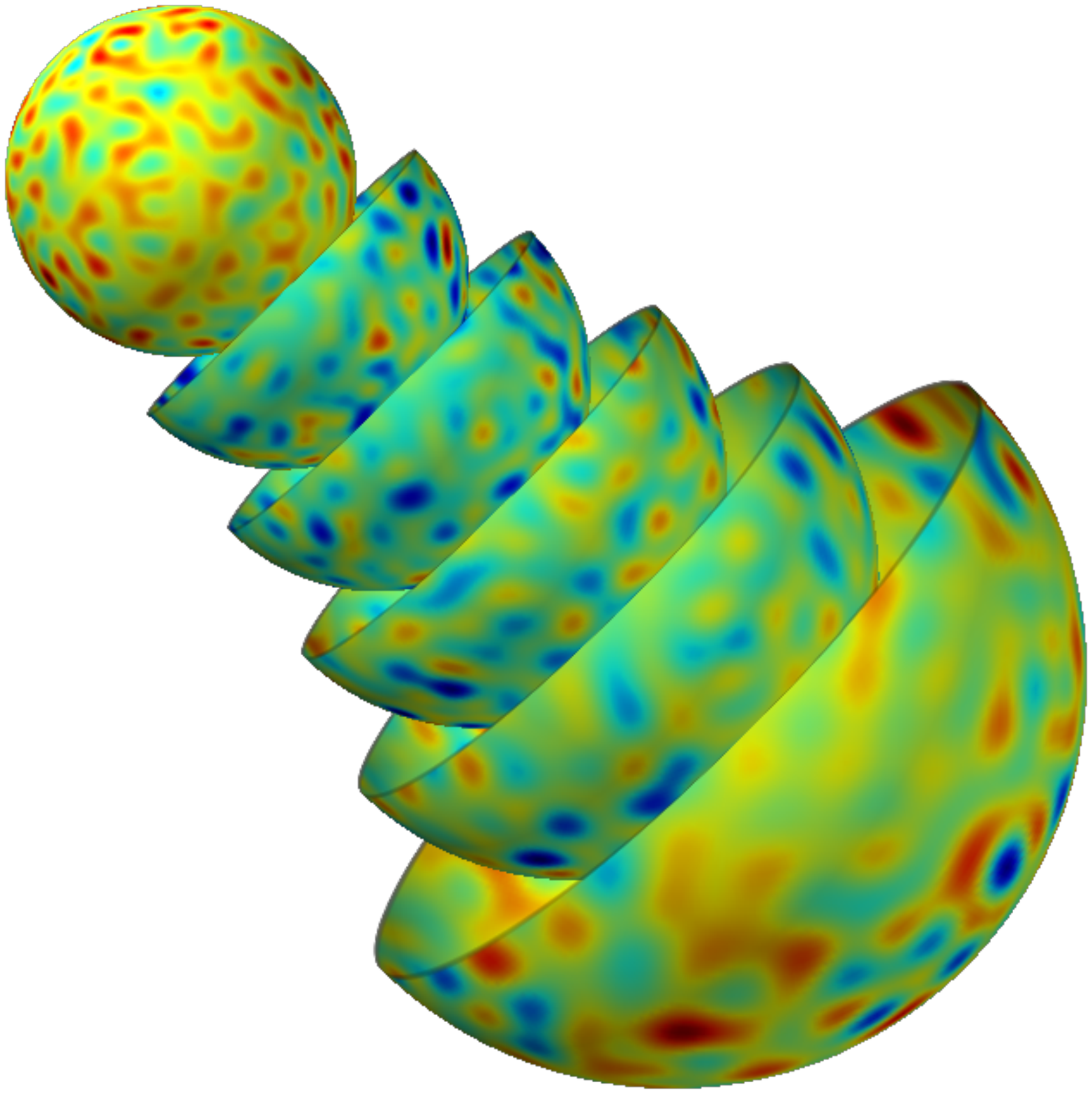}}
    \put(-72,63){$\bm{x}^{\text{DIR}}$}
    \put(-92, -7){\footnotesize SNR = $-2.707$ dB}
  \end{minipage}\\
  \vspace{-10pt}
  \begin{minipage}{0.38\linewidth}
    \centering \centering \includegraphics[width=\linewidth, trim={5.3cm 2cm 5.3cm 0cm}, clip]{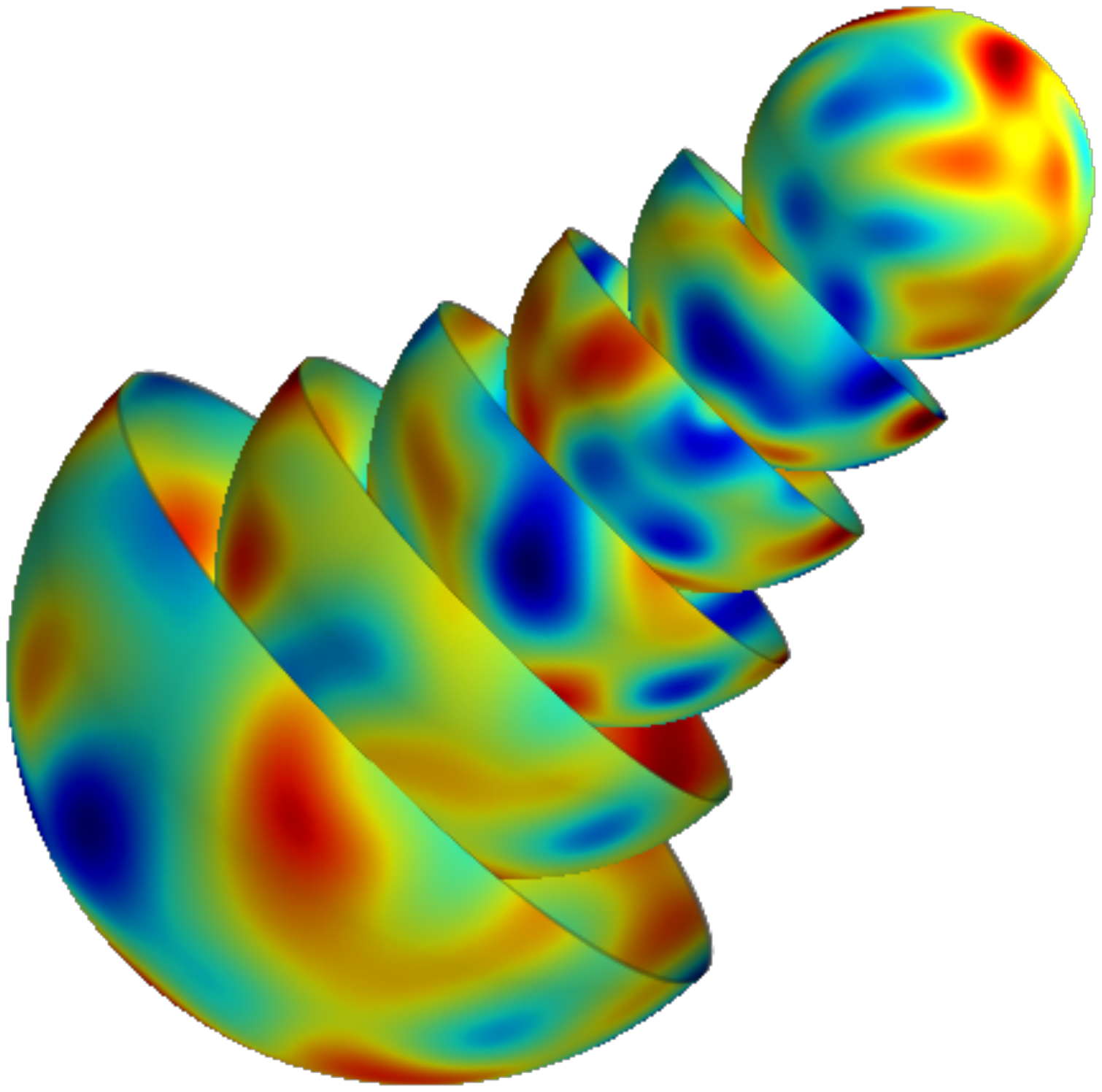}
    \put(-72,66){$\bm{x}^{\text{MAP}}$}
    \put(-92, -4){\footnotesize SNR = $10.293$ dB}
  \end{minipage}
  \begin{minipage}{0.38\linewidth}
    \centering \centering \includegraphics[width=\linewidth, trim={6cm 2cm 6cm 0cm}, clip]{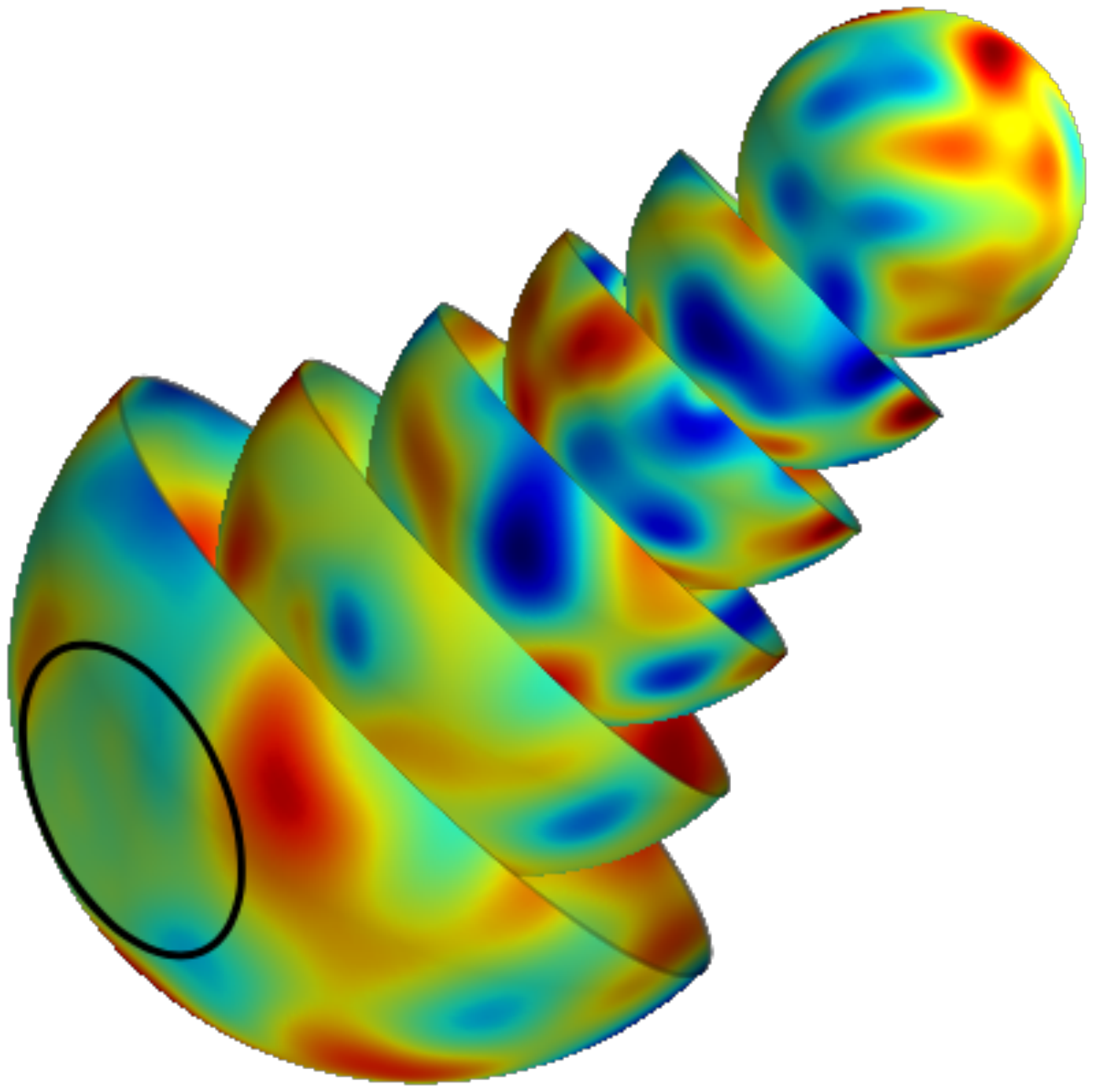}
    \put(-72,67){$\bm{x}^{\text{SUR}}$}
    \put(-98, 0){\footnotesize Bayesian Hypothesis Test}
    \put(-83,40){$\Omega$}
  \end{minipage}

  \caption{\textbf{Description:} Variational inference results for 3D directional deconvolution
    with inpainting using pseudo-Gaussian simulations for $L=P=32$ (upsampled to 128), roughly approximating physical fields \emph{e.g.} atmospheric fields -- these
    methods trivially extend to realistic simulations. Specifically the ground
    truth is smoothed with a directional kernel, 50\% of observations are masked, and the remaining
    observations are corrupted with 30dB \emph{i.i.d.} Gaussian noise \mbox{$\bm{n} \sim \mathcal{N}(0, \sigma^2)$}. \textbf{Panels:} Input
    ground truth (top left), naive inversion (baseline, top right), \emph{maximum a posteriori} (MAP) estimator
    using ball wavelet $\ell_1$ sparsity (bottom left), and Bayesian hypothesis test of local structure $\Omega$ (bottom right, see Section \ref{sec:UQ}).
    \textbf{Discussion:} Notice that naive direct inversion recovers a poor estimator (SNR = $-2.707$dB). Alternatively, treating the problem as a Bayesian
    variational problem not only recovers a very good estimate (SNR = $10.293$dB), but also
    supports principled uncertainty quantification. The Bayesian hypothesis test of local substructure $\Omega$ in the bottom right subfigure correctly determines the
    physicality of this feature at $99\%$-confidence. The MAP estimate and uncertainties were recovered in $\sim 2$ minutes of non-dedicated
    compute on a MacBook Air 2016 respectively, which can trivially sped up through \emph{e.g.} OpenMP and MPI.
  }
  \label{fig:ball_plot}
\end{figure}

\section{Numerical experiment} \label{sec:examples}
In this section we consider a noisy inpainting directional deconvolution inverse problem, which is (seriously) ill-posed and ill-conditioned. Such an example is representative of a diverse set of practical applications. Consider again the problem setup outlined in Section \ref{sec:bayesian_optimisation}, where we model the acquisition
of observations by the sensing operator
\begin{equation}
  \PHI = \bm{\mathsf{M}} \bm{\mathsf{B}}^{-1} \bm{\mathsf{K}} \bm{\mathsf{B}}
  \quad \text{and} \quad
  \PHI^\dagger = \bm{\mathsf{B}}^\dagger \bm{\mathsf{K}} \bm{\mathsf{B}}^{-\dagger} \bm{\mathsf{M}}^\dagger,
\end{equation}
where $\bm{\mathsf{B}}$ and $\bm{\mathsf{B}}^{-1}$ represent forward and inverse spin-$0$
Spherical-Laguerre transforms (see Section \ref{sec:spin_signals}), $\bm{\mathsf{K}}$ is
multiplication with a skewed Gaussian kernel in Spherical-Laguerre
space (which is trivially self-adjoint), $\bm{\mathsf{M}}$ represents masking, and $\dagger$ denotes the operator
adjoint. It is important to note that \mbox{$\bm{\mathsf{B}}^{-1} \not= \bm{\mathsf{B}}^\dagger$} which is a
poorly motivated approximation often adopted in settings involving spherical harmonic
transforms. Additionally, we define as a baseline the naive direct inversion
\mbox{$\bm{x}^{\text{DIR}} =  \PHI^{-1} \bm{y}$} for \mbox{$\PHI^{-1} =\bm{\mathsf{B}}^{-1} \bm{\mathsf{K}}^{-1}  \bm{\mathsf{B}} \bm{\mathsf{M}}^{\dagger}$}, where $\bm{\mathsf{K}}^{-1}$ is simply division by the Spherical-Laguerre space convolutional kernel. As we are considering ill-posed inverse problems \cite{hadamard:1902} the naive inverse \mbox{$\bm{x}^{\text{DIR}}$} can give (potentially non-physical) solutions which lie far from the true signal. Moreover, the noise contribution, which is typically highly oscillatory, may (and often does) dominate the solution.

We consider the case in which $\bm{n}$ is independent and identically distributed noise drawn
from a univariate Gaussian distribution \mbox{$\bm{n} \sim \mathcal{N}(0, \sigma^2)$}. Our likelihood function
is thus given by a Gaussian distribution with zero mean and variance $\sigma^2$. Suppose
our prior knowledge indicates that $\bm{x}$ is likely to be sparsely distributed when projected
into the ball wavelet dictionary $\bm{\mathsf{\Psi}}$, described in Section \ref{sec:ball_wavelets}. A prior
distribution which naturally promotes sparsity is the Laplacian distribution, which one might
adopt, such that the posterior is given by
\begin{equation}
  P(\bm{x} | \bm{y}; \PHI) \propto \exp \bigg ( \frac{-\bnorm{\PHI\bm{x} - \bm{y}}_2^2}{2\sigma^2} \bigg )
  \exp \big ( -\lambda \pnorm{\bm{\mathsf{\Psi}}^\dagger\bm{x}}_1 \big ),
\end{equation}
where $\bnorm{\cdot}$ and $\pnorm{\cdot}$ are the standard $\ell_p$-norms weighted by
pixel-size so as to better approximate the continuous $\ell_p$-norms on the ball. By following the logic presented in Section \ref{sec:bayesian_optimisation}
one finds the MAP estimate is given by
\begin{equation}
  \bm{x}^{\text{MAP}} = \argminT_{\bm{x} \in \mathbb{R}^{N_{\Bt}}} \bigg \lbrace
  \frac{\bnorm{\PHI \bm{x} - \bm{y}}_2^2}{2\sigma^2} + \lambda\pnorm{\bm{\mathsf{\Psi}}^\dagger\bm{x}}_1 \bigg \rbrace,
\end{equation}
with regularization parameter \mbox{$\lambda \in \Rp$} which we marginalize over \cite{pereyra:2015}, to maintain a principled Bayesian interpretation.

\subsection{Experiment details}
We generate a ground truth signal $\bm{x}^{\text{TRUE}}$ by smoothing a random
signal on the ball, effectively generating a pseudo-Gaussian random field, which is bandlimited
at $L$ in the angular domain and $P=L$ along the radial line. This ground truth is mapped by $\PHI$ to simulated observations which are subsequently polluted with \emph{i.i.d.} noise, drawn from a univariate Gaussian
distribution, to form simulated observations $\bm{y}$, such that the input signal to noise ratio,
\begin{equation}
  \text{SNR} = 20 \times \log_{10}\bigg (\frac{\norm{\PHI \bm{x}^{\text{TRUE}}}_2}{\norm{\PHI \bm{x}^{\text{TRUE}} - \bm{y}}_2} \bigg ),
\end{equation}
is 30dB. An analogous SNR definition is used to quantify the reconstruction fidelity between $\bm{x}$ and a recovered solution $\bm{x}^*$. Both the naive inversion (SNR$=-2.707$dB), and MAP (SNR$=10.293$dB) estimators are recovered, and are presented in Figure \ref{fig:ball_plot}. 
Note that the variational solution is recovered in the analysis unconstrained setting through the proximal forward-backward algorithm \cite{beck:2009, combettes:2011}. 
This dramatic improvement in reconstruction fidelity is compounded by the fact that our estimator also supports principled Bayesian uncertainty quantification, namely hypothesis testing of structure \emph{e.g.} the diffuse, high intensity region $\Omega$ 
highlighted in Figure \ref{fig:ball_plot} was correctly determined to be physical at $99\%$ confidence.

\section{Discussion \& conclusions} \label{sec:conclusions}
Whilst there are many methods which consider reconstruction over the 3D ball by analyzing
individual concatenated spherical shells, to the best of our knowledge,
this is the first article which develops variational regularization methods natively on the ball.
Leveraging recent developments in probability concentration theory, we demonstrate how
MAP estimation (unconstrained optimization) permits principled uncertainty quantification. Our Bayesian variational approach benefits from the computational efficiency
of convex optimization whilst facilitating principled uncertainty quantification.
We demonstrate that our variational approach is effective at solving seriously
ill-posed and ill-conditioned inverse problems on the ball, recovering very accurate, robust estimates of
the underlying ground truth.
In future collaborative work we will
apply these methods to more realistic simulations and observational data, for a variety of application domains. As a bi-product
of this work an open-source, flexible, scalable object oriented C++ software,
B3INV\footnote{\url{https://github.com/astro-informatics/b3inv}} was created which is constructed on the convex optimization package SOPT\footnote{\url{http://astro-informatics.github.io/sopt/}} \cite{carrillo:2012:sara, carrillo:2013:sara, onose:2016:sopt, pratley:2018:sopt}. Additionally, fast adjoint operators which were written and collected into the FLAG and FLAGLET codebases.

\ifCLASSOPTIONcaptionsoff
  \newpage
\fi


\begin{thebibliography}{10}
\providecommand{\url}[1]{#1}
\csname url@rmstyle\endcsname
\providecommand{\newblock}{\relax}
\providecommand{\bibinfo}[2]{#2}
\providecommand\BIBentrySTDinterwordspacing{\spaceskip=0pt\relax}
\providecommand\BIBentryALTinterwordstretchfactor{4}
\providecommand\BIBentryALTinterwordspacing{\spaceskip=\fontdimen2\font plus
\BIBentryALTinterwordstretchfactor\fontdimen3\font minus
  \fontdimen4\font\relax}
\providecommand\BIBforeignlanguage[2]{{%
\expandafter\ifx\csname l@#1\endcsname\relax
\typeout{** WARNING: IEEEtran.bst: No hyphenation pattern has been}%
\typeout{** loaded for the language `#1'. Using the pattern for}%
\typeout{** the default language instead.}%
\else
\language=\csname l@#1\endcsname
\fi
#2}}

\bibitem{tuch:2004}
D.~S. Tuch, ``Q-ball imaging,'' \emph{Magnetic Resonance in Medicine: An
  Official Journal of the International Society for Magnetic Resonance in
  Medicine}, vol.~52, no.~6, pp. 1358--1372, 2004.

\bibitem{simons:2011}
F.~J. {Simons} \emph{et~al.}, ``{Solving or resolving global tomographic models
  with spherical wavelets, and the scale and sparsity of seismic
  heterogeneity},'' \emph{Geophysical Journal International}, vol. 187, no.~2,
  pp. 969--988, Nov. 2011.

\bibitem{marignier:2020}
A.~Marignier, A.~M.~G. Ferreira, and T.~Kitching, ``The probability of mantle
  plumes in global tomographic models,'' \emph{Geochemistry, Geophysics,
  Geosystems}, vol.~21, no.~9, p. e2020GC009276, 2020.

\bibitem{kendall:2021}
E.~Kendall, A.~Ferreira, S.-J. Chang, M.~Witek, and D.~Peter, ``Constraints on
  the upper mantle structure beneath the pacific from 3-d anisotropic waveform
  modelling,'' \emph{Journal of Geophysical Research: Solid Earth}.

\bibitem{heavens:2003}
A.~Heavens, ``{3D weak lensing},'' \emph{MNRAS}, vol. 343, no.~4, pp.
  1327--1334, 08 2003.

\bibitem{leistedt:2015}
B.~{Leistedt}, J.~D. {McEwen}, T.~D. {Kitching}, and H.~V. {Peiris}, ``{3D weak
  lensing with spin wavelets on the ball},'' \emph{Physical Review D}, vol.~92,
  no.~12, p. 123010, Dec. 2015.

\bibitem{boomsma:2017}
W.~Boomsma and J.~Frellsen, ``Spherical convolutions and their application in
  molecular modelling.'' in \emph{Advances in Neural Information Processing
  Systems}, vol.~2, 2017, p.~6.

\bibitem{mcewen:2013:sparse}
J.~D. {McEwen}, G.~{Puy}, J.~{Thiran}, P.~{Vandergheynst}, D.~{Van De Ville},
  and Y.~{Wiaux}, ``Sparse image reconstruction on the sphere: Implications of
  a new sampling theorem,'' \emph{IEEE Transactions on Image Processing},
  vol.~22, no.~6, pp. 2275--2285, 2013.

\bibitem{wallis:2017:sparse}
C.~G.~R. {Wallis}, Y.~{Wiaux}, and J.~D. {McEwen}, ``{Sparse Image
  Reconstruction on the Sphere: Analysis and Synthesis},'' \emph{IEEE
  Transactions on Image Processing}, vol.~26, pp. 5176--5187, Nov. 2017.

\bibitem{price:2021:s2inv}
M.~A. {Price}, L.~{Pratley}, and J.~D. {McEwen}, ``{Sparse image reconstruction
  on the sphere: a general approach with uncertainty quantification},''
  \emph{submitted to IEEE Transactions on Image Processing}, p.
  arXiv:2105.04935, 5 2021.

\bibitem{donoho:2006:compressed}
D.~L. Donoho, ``Compressed sensing,'' \emph{IEEE Transactions on information
  theory}, vol.~52, no.~4, pp. 1289--1306, 2006.

\bibitem{candes:2006:compressive}
E.~J. Cand{\`e}s \emph{et~al.}, ``Compressive sampling,'' 2006.

\bibitem{chan:2017:curvelets}
J.~Y.~H. {Chan}, B.~{Leistedt}, T.~D. {Kitching}, and J.~D. {McEwen},
  ``Second-generation curvelets on the sphere,'' \emph{IEEE Transactions on
  Signal Processing}, vol.~65, no.~1, pp. 5--14, 2017.

\bibitem{leistedt:2013:s2let}
{Leistedt, B.}, {McEwen, J. D.}, {Vandergheynst, P.}, and {Wiaux, Y.}, ``S2let:
  A code to perform fast wavelet analysis on the sphere,'' \emph{A\&A}, vol.
  558, p. A128, 2013.

\bibitem{mcewen:2019:ridgelets}
J.~D. {McEwen} and M.~A. {Price}, ``Scale-discretised ridgelet transform on the
  sphere,'' in \emph{2019 27th European Signal Processing Conference
  (EUSIPCO)}, 2019, pp. 1--5.

\bibitem{michel:2005}
V.~Michel, ``Wavelets on the 3 dimensional ball,'' \emph{PAMM}, vol.~5, no.~1,
  pp. 775--776, 2005.

\bibitem{fengler:2006}
M.~Fengler, D.~Michel, and V.~Michel, \emph{ZAMM - Journal of Applied
  Mathematics and Mechanics}, vol.~86, no.~11, pp. 856--873, 2006.

\bibitem{lanusse:2012}
F.~{Lanusse}, A.~{Rassat}, and J.~L. {Starck}, ``{Spherical 3D isotropic
  wavelets},'' \emph{Journal of Astronomy \& Astrophysics}, vol. 540, p. A92,
  Apr. 2012.

\bibitem{durastanti:2014}
C.~Durastanti, Y.~Fantaye, F.~Hansen, D.~Marinucci, and I.~Z. Pesenson,
  ``Simple proposal for radial 3d needlets,'' \emph{Phys. Rev. D}, vol.~90, p.
  103532, Nov 2014.

\bibitem{khalid:2016}
Z.~Khalid, R.~A. Kennedy, and J.~D. McEwen, ``Slepian spatial-spectral
  concentration on the ball,'' \emph{Applied and Computational Harmonic
  Analysis}, vol.~40, no.~3, pp. 470--504, 2016.

\bibitem{leistedt:2012}
B.~{Leistedt} and J.~D. {McEwen}, ``{Exact Wavelets on the Ball},'' \emph{IEEE
  Transactions on Signal Processing}, vol.~60, no.~12, pp. 6257--6269, Dec.
  2012.

\bibitem{pereyra:2017}
M.~Pereyra, ``Maximum-a-posteriori estimation with bayesian confidence
  regions,'' \emph{SIAM Journal on Imaging Sciences}, vol.~10, no.~1, pp.
  285--302, 2017.

\bibitem{pollard:1947}
H.~Pollard, ``Representation of an analytic function by a laguerre series,''
  \emph{Annals of Mathematics}, vol.~48, no.~2, pp. 358--365, 1947.

\bibitem{weniger:2008}
E.~J. {Weniger}, ``{On the analyticity of Laguerre series},'' \emph{Journal of
  Physics A Mathematical General}, vol.~41, no.~42, p. 425207, Oct. 2008.

\bibitem{newman:1966}
E.~T. Newman and R.~Penrose, ``Note on the bondi-metzner-sachs group,''
  \emph{Journal of Mathematical Physics}, vol.~7, no.~5, pp. 863--870, 1966.

\bibitem{goldberg:1967}
J.~N. Goldberg, A.~J. Macfarlane, E.~T. Newman, F.~Rohrlich, and E.~C.~G.
  Sudarshan, ``Spin-s spherical harmonics,'' \emph{Journal of Mathematical
  Physics}, vol.~8, no.~11, pp. 2155--2161, 1967.

\bibitem{mcewen:2011:ssht}
J.~D. {McEwen} and Y.~{Wiaux}, ``{A Novel Sampling Theorem on the Sphere},''
  \emph{IEEE Transactions on Signal Processing}, vol.~59, no.~12, pp.
  5876--5887, Dec. 2011.

\bibitem{mcewen:2015:s2let}
J.~D. {McEwen}, B.~{Leistedt}, M.~{B{\"u}ttner}, H.~V. {Peiris}, and
  Y.~{Wiaux}, ``{Directional spin wavelets on the sphere},'' \emph{arXiv
  e-prints}, p. arXiv:1509.06749, Sept. 2015.

\bibitem{cai:2018:uq}
X.~{Cai}, M.~{Pereyra}, and J.~D. {McEwen}, ``{Uncertainty quantification for
  radio interferometric imaging: II. MAP estimation},'' \emph{MNRAS}, vol. 480,
  no.~3, pp. 4170--4182, Nov. 2018.

\bibitem{price:2018}
M.~A. {Price}, X.~{Cai}, J.~D. {McEwen}, T.~D. {Kitching}, and C.~G.~R.
  {Wallis}, ``{Sparse Bayesian mass-mapping with uncertainties: hypothesis
  testing of structure},'' \emph{submitted to MNRAS}, 2018.

\bibitem{price:2021:darkmapper}
M.~A. {Price}, J.~D. {McEwen}, L.~{Pratley}, and T.~D. {Kitching}, ``{Sparse
  Bayesian mass-mapping with uncertainties: Full sky observations on the
  celestial sphere},'' \emph{MNRAS}, vol. 500, no.~4, pp. 5436--5452, Jan.
  2021.

\bibitem{press:2007}
W.~H. Press, S.~A. Teukolsky, W.~T. Vetterling, and B.~P. Flannery,
  \emph{Numerical Recipes 3rd Edition: The Art of Scientific Computing},
  3rd~ed.\hskip 1em plus 0.5em minus 0.4em\relax USA: Cambridge University
  Press, 2007.

\bibitem{mcewen:2013:flaglet}
J.~D. {McEwen} and B.~{Leistedt}, ``{Fourier-Laguerre transform, convolution
  and wavelets on the ball},'' \emph{arXiv e-prints}, p. arXiv:1307.1307, July
  2013.

\bibitem{wiaux:2008}
Y.~Wiaux, J.~D. McEwen, P.~Vandergheynst, and O.~Blanc, ``{Exact reconstruction
  with directional wavelets on the sphere},'' \emph{MNRAS}, vol. 388, no.~2,
  pp. 770--788, 07 2008.

\bibitem{mcewen:2018:localisation}
J.~D. McEwen, C.~Durastanti, and Y.~Wiaux, ``Localisation of directional
  scale-discretised wavelets on the sphere,'' \emph{Applied and Computational
  Harmonic Analysis}, vol.~44, no.~1, pp. 59--88, 2018.

\bibitem{hadamard:1902}
J.~Hadamard, ``Sur les probl{\`e}mes aux d{\'e}riv{\'e}es partielles et leur
  signification physique,'' \emph{Princeton university bulletin}, pp. 49--52,
  1902.

\bibitem{combettes:2011}
P.~L. Combettes and J.-C. Pesquet, ``Proximal splitting methods in signal
  processing,'' in \emph{Fixed-point algorithms for inverse problems in science
  and engineering}.\hskip 1em plus 0.5em minus 0.4em\relax Springer, 2011, pp.
  185--212.

\bibitem{moreau:1962}
J.~J. Moreau, ``Fonctions convexes duales et points proximaux dans un espace
  hilbertien,'' \emph{C.R. Acad. Sci. Paris Ser. A Math.}, vol. 255, pp.
  2897--2899, 1962.

\bibitem{boyd:2011:admm}
S.~Boyd, N.~Parikh, and E.~Chu, \emph{Distributed optimization and statistical
  learning via the alternating direction method of multipliers}.\hskip 1em plus
  0.5em minus 0.4em\relax Now Publishers Inc, 2011.

\bibitem{price:2019:lci}
M.~A. {Price}, X.~{Cai}, J.~D. {McEwen}, M.~{Pereyra}, and T.~D. {Kitching},
  ``{Sparse Bayesian mass mapping with uncertainties: local credible
  intervals},'' \emph{MNRAS}, vol. 492, no.~1, pp. 394--404, Dec. 2019.

\bibitem{pereyra:2016:proximal}
M.~Pereyra, ``Proximal markov chain monte carlo algorithms,'' \emph{Statistics
  and Computing}, vol.~26, no.~4, pp. 745--760, 2016.

\bibitem{repetti:2018}
A.~Repetti, M.~Pereyra, and Y.~Wiaux, ``Scalable bayesian uncertainty
  quantification in imaging inverse problems via convex optimization,''
  \emph{SIAM Journal on Imaging Sciences}, vol.~12, no.~1, pp. 87--118, 2019.

\bibitem{price:2019:peaks}
M.~A. Price, J.~D. McEwen, X.~Cai, and T.~D. Kitching, ``{Sparse Bayesian mass
  mapping with uncertainties: peak statistics and feature locations},''
  \emph{MNRAS}, vol. 489, no.~3, pp. 3236--3250, 08 2019.

\bibitem{pereyra:2015}
M.~Pereyra, J.~Bioucas-Dias, and M.~Figueiredo, \emph{Maximum-a-posteriori
  estimation with unknown regularisation parameters}, 12 2015, pp. 230--234.

\bibitem{beck:2009}
A.~Beck and M.~Teboulle, ``A fast iterative shrinkage-thresholding algorithm
  for linear inverse problems,'' \emph{SIAM journal on imaging sciences},
  vol.~2, no.~1, pp. 183--202, 2009.

\bibitem{carrillo:2012:sara}
R.~E. {Carrillo}, J.~D. {McEwen}, and Y.~{Wiaux}, ``{Sparsity Averaging
  Reweighted Analysis (SARA): a novel algorithm for radio-interferometric
  imaging},'' \emph{MNRAS}, vol. 426, no.~2, pp. 1223--1234, Oct. 2012.

\bibitem{carrillo:2013:sara}
R.~E. {Carrillo}, J.~D. {McEwen}, D.~{Van De Ville}, J.-P. {Thiran}, and
  Y.~{Wiaux}, ``{Sparsity Averaging for Compressive Imaging},'' \emph{IEEE
  Signal Processing Letters}, vol.~20, no.~6, pp. 591--594, June 2013.

\bibitem{onose:2016:sopt}
A.~{Onose}, R.~E. {Carrillo}, A.~{Repetti}, J.~D. {McEwen}, J.-P. {Thiran},
  J.-C. {Pesquet}, and Y.~{Wiaux}, ``{Scalable splitting algorithms for
  big-data interferometric imaging in the SKA era},'' \emph{MNRAS}, vol. 462,
  no.~4, pp. 4314--4335, Nov. 2016.

\bibitem{pratley:2018:sopt}
L.~{Pratley}, J.~D. {McEwen}, M.~{d'Avezac}, R.~E. {Carrillo}, A.~{Onose}, and
  Y.~{Wiaux}, ``{Robust sparse image reconstruction of radio interferometric
  observations with PURIFY},'' \emph{MNRAS}, vol. 473, no.~1, pp. 1038--1058,
  Jan. 2018.

\end{thebibliography}

\vfill
\end{document}